\documentclass[12pt]{article}

\usepackage{graphicx,bm,amssymb,amsmath}

\newcommand{\bi}{\begin{itemize}}
\newcommand{\ei}{\end{itemize}}
\newcommand{\ben}{\begin{enumerate}}
\newcommand{\een}{\end{enumerate}}
\newcommand{\be}{\begin{equation}}
\newcommand{\ee}{\end{equation}}
\newcommand{\bea}{\begin{eqnarray}} 
\newcommand{\eea}{\end{eqnarray}}
\newcommand{\bc}{\begin{center}}
\newcommand{\ec}{\end{center}}

\newcommand{\eg}{{\it e.g.\ }}

\newcommand{\tbox}[1]{{\mbox{\tiny #1}}}
\newcommand{\mbf}[1]{{\mathbf #1}}
\newcommand{\ofr}{(\mbf{r})}
\newcommand{\ofs}{(\mbf{s})}
\newcommand{\vO}{\mbox{vol($\Omega$)}}
\newcommand{\ep}{\epsilon}
\newcommand{\et}{{\cal E}}
\newcommand{\epij}{{\epsilon_{ij}}}
\newcommand{\etij}{{{\cal E}_{ij}}}
\newcommand{\rruv}{{\langle u, r^2v \rangle_\Omega}}
\newcommand{\rrij}{{\langle \phi_i, r^2\phi_j \rangle_\Omega}}
\newcommand{\rrii}{{\langle \phi_i, r^2\phi_i \rangle_\Omega}}
\newcommand{\rrjj}{{\langle \phi_j, r^2\phi_j \rangle_\Omega}}
\newcommand{\uv}{{\langle u, v \rangle_\Omega}}
\newcommand{\ij}{{\langle \phi_i, \phi_j \rangle_\Omega}}
\newcommand{\ds}{{dA}}
\newcommand{\dr}{{dV}}
\newcommand{\np}{{N_{\mbf{p}}}}
\newcommand{\nq}{{N_q}}
\newcommand{\mku}{-E_u} 
\newcommand{\mkv}{-E_v}
\newcommand{\di}[2]{\ensuremath{\overset{\mbox{\raisebox{-.3ex}[.7ex][0ex]%
	{\ensuremath{#2}}}}{#1}}} 
\newcommand{\lc}{\ensuremath{\scriptstyle \smallfrown}} 
\newcommand{\lla}{\ensuremath{\,\scriptstyle \smallfrown%
	\mkern-16.5mu\displaystyle \smallfrown}}    
\newcommand{\da}[1]{\di{#1}{.}}      
\newcommand{\la}[1]{\di{#1}{\lc}}      
\newcommand{\lr}{\ensuremath{\leftrightarrow}}
\newcommand{\ls}{\ensuremath{\scriptscriptstyle \cap}} 
\newtheorem{thm}{Theorem}

\newtheorem{lem}{Lemma}[thm]
\newtheorem{cor}{Corollary}[thm]

\title{Quasi-orthogonality on the boundary for Euclidean Laplace eigenfunctions}
\author{Alex H. Barnett\\
{\normalsize \em Courant Institute, New York University,
251 Mercer St, New York, NY 10012}\\ {\small \tt barnett@cims.nyu.edu}}
\date{July 3, 2004}

\begin{document}
\maketitle

\begin{abstract}
Consider the Laplacian
in a bounded domain in $\mathbb{R}^d$
with general (mixed) homogeneous boundary conditions.
We prove that its eigenfunctions
are `quasi-orthogonal' on the boundary with respect to a certain norm.
%
Boundary orthogonality is proved asymptotically
within a narrow eigenvalue window of width $o(E^{1/2})$
centered about $E$, as $E\rightarrow\infty$.
For the special case of Dirichlet boundary conditions,
the normal-derivative functions are quasi-orthogonal
on the boundary with respect to the geometric
weight function $\mbf{r}\cdot\mbf{n}$.
%
%
The result is independent of any quantum ergodicity assumptions
and hence of the nature of
the domain's geodesic flow;
however if this is ergodic
then heuristic semiclassical results
suggest an improved asymptotic estimate.
Boundary quasi-orthogonality
is the key to a highly efficient `scaling method' for
numerical solution of the Laplace eigenproblem at large eigenvalue.
One of the main results of this paper is then to place this method
on a more rigorous footing.
\end{abstract}

\section{Introduction and main results}
\label{sec:main}

Let $\Omega\subset\mathbb{R}^d$ be an open bounded Euclidean domain of
dimension $d\ge 2$, with boundary $\Gamma\equiv\partial\Omega$
parametrized by the $(d{-}1)$ dimensional
coordinate $\mbf{s}\in\Gamma$,
and outwards unit normal vector $\mbf{n}(\mbf{s})$.
For instance we envisage the boundary being a finite union of smooth
surfaces with angles at junctions bounded away from zero and
$2\pi$;
in general our domain can be Lipschitz.
The set of eigenfunctions $\{\phi_i\ofr\}$, $i=1,2\ldots\infty$
of the Laplace operator $\Delta$
are defined by
\be
	-\Delta \phi_i \;= \; E_i \phi_i,
\label{eq:helm}
\ee
everywhere in $\Omega$,
and by Dirichlet boundary conditions on some subset of the boundary,
\be
	\phi_i\ofs \; =  \; 0
		\qquad \mbox{for } \mbf{s}\in\Gamma_D,
\label{eq:dbcs}
\ee
and homogeneous boundary conditions on the remainder,
\be
	\gamma\ofs\phi_i\ofs + \partial_n\phi_i\ofs\;=\;0
		\qquad \mbox{for } \mbf{s}\in\Gamma\setminus \Gamma_D, 
\label{eq:bcs}
\ee
where $\partial_n\phi_i\ofs \equiv \mbf{n}\ofs\cdot\nabla\phi_i\ofs$
is the normal derivative at the boundary.
We assume the subset $\Gamma_D \subseteq \Gamma$ is composed of a finite number
of compact pieces, and that on the remainder $\gamma\ofs$ is bounded.
The boundary conditions are such that $\Delta$ is
self-adjoint, and we also assume
they are such that a complete set of eigenfunctions exists.
The corresponding eigenvalues
are ordered $E_1 \le E_2 \le \cdots$, and
eigenfunctions are orthonormal on the domain, $\ij = \delta_{ij}$.
In keeping with common quantum-mechanical
terminology we will say that level $i$ has energy $E_i$.

Let $q$ be the following symmetric bilinear form
depending on an energy parameter $\et$,
\be
	q(u,v; \et) \; := \; \oint_{\Gamma}\!
	\frac{d-2}{2}(u_n v+v_n u)
	+ r_n\left(\frac{\et}{2} uv - \nabla u \cdot \nabla v\right)
	+ u_r v_n + v_r u_n \, \ds.
\label{eq:qdef}
\ee
Its particular form has properties that will become apparent shortly.
Here $u$ and $v$ are functions defined with their derivatives
on $\overline{\Omega}$, the closure of $\Omega$.
We use
the abbreviations $u_n \equiv \partial_n u$,
$u_r \equiv \partial_r u \equiv \mbf{r} \cdot \nabla u$,
and $r_n \equiv \mbf{r}\cdot\mbf{n}$.
The surface element on $\Gamma$ is $\ds$.
Note that $q$ is
sensitive only to boundary values and first
derivatives of $u$ and $v$.

Let $Q$ be the semi-infinite symmetric matrix
defined using the eigenfunctions of domain $\Omega$ by
\be
	Q_{ij} \; := \; q(\phi_i, \phi_j; \etij), \qquad i,j\ge 1
\label{eq:q}
\ee
where $\etij\equiv E_i+E_j$ is the sum of the energy eigenvalues.

Our main result is
\begin{thm}[Quasi-orthogonality] 
The matrix $Q$ defined above has diagonal elements
\be
	Q_{ii} \; = \; 2E_i, \qquad \mbox{for all } i \ge 1,
\label{eq:thm1diag}
\ee
and off-diagonal elements satisfying
\be
	|Q_{ij}| \; \le \; C(E_i-E_j)^2, \qquad \mbox{for all } i\neq j,
	\quad i,j\ge 1,
\label{eq:bound}
\ee
where the constant $C$ depends only on the shape of the domain.
\label{thm:q}
\end{thm} 

Note that by (\ref{eq:thm1diag}) and (\ref{eq:bound})
the diagonal elements grow linearly with energy whereas off-diagonal
elements are bounded by a fixed quadratic function of
the energy difference.

Bearing in mind that eigenfunction derivatives grow like
$\|\nabla \phi_i\|_{L^2(\Omega)} = O(E_i^{1/2})$,
we might naively expect from the size of terms in (\ref{eq:qdef}) that
all off-diagonal elements of $Q_{ij}$ near the diagonal ($E_i\approx E_j$)
grow linearly with $E_i$, as $E_i$ (and therefore $E_j$) tends to infinity.
Theorem~\ref{thm:q} tells us that this is not so, and that the choice of
$q$ carries with it non-trivial cancellation.
This particular property of $q$ motivates our
choice of the bilinear form.
We conjecture that this
property is {\em unique} to the form $q$ given by (\ref{eq:qdef}),
up to a choice of origin.
It should be emphasized that this result holds for all
elements $i,j$, not just asymptotically.
We immediately have the following
\begin{cor}[Asymptotic orthogonality] 
Given $Q$ as defined above, let $\tilde{Q}$ be the sub-matrix of $Q$
corresponding to all levels $i,j$ in a local energy window
$E_i, E_j \in [E-\ep_0, E+\ep_0]$ where $\ep_0 = O(E^\beta)$.
Then for all $\beta<1/2$,
\ben
\item
\be
	\frac{\tilde{Q}}{2E} \rightarrow I
	\qquad\mbox{as}\quad E \rightarrow \infty,
\ee
where $I$ is the identity matrix.
Thus, close to the diagonal,
boundary orthogonality is asymptotically exact.
\item
Off-diagonal elements of $\tilde{Q}/2E$ are
bounded in absolute value by $c_\beta E^{2\beta-1}$.
\een
\label{cor:q}
\end{cor} 

\begin{figure}
\bc
\includegraphics[width=\textwidth]{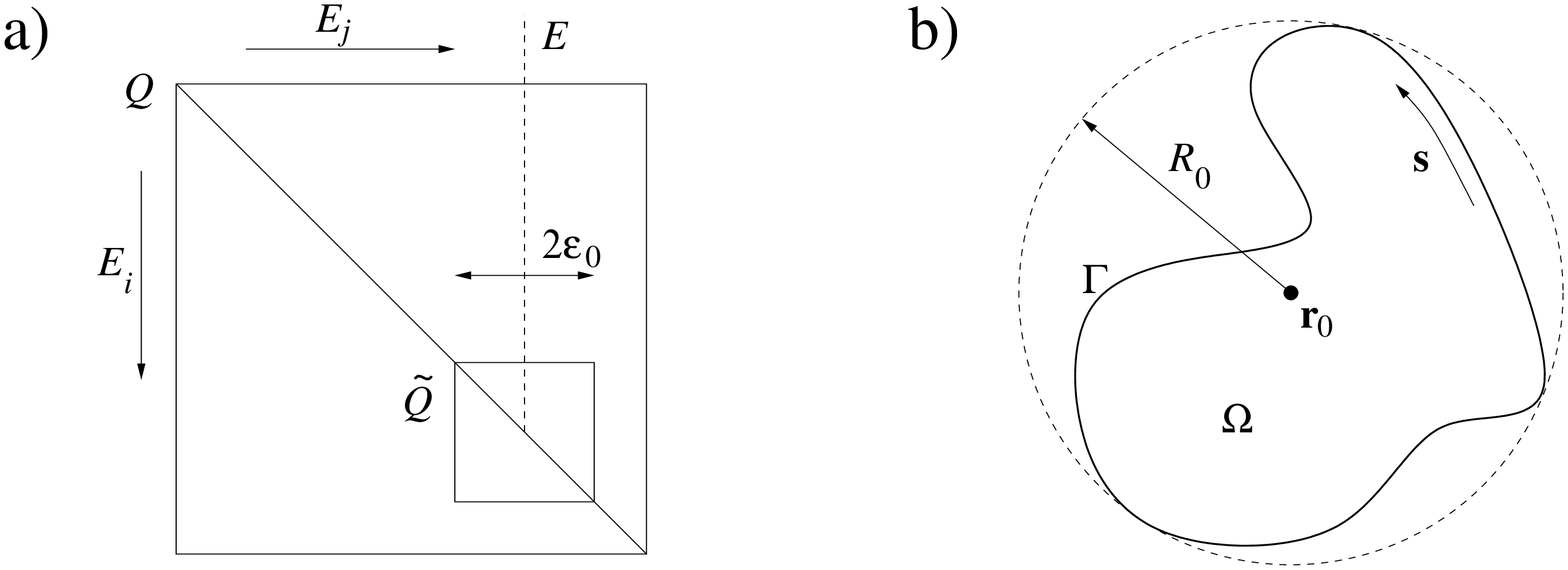}
\ec
\caption{
a) Illustrates the extraction of $\tilde{Q}$ from the full
matrix $Q$.
b) Illustrates (for $d{=}2$) the geometry of the domain, and its
escribed circle.
\label{fig}
}
\end{figure}

The sub-matrix $\tilde{Q}$ is positioned in $Q$ as indicated in
Fig.~\ref{fig}a.
Note that by Weyl's law for the asymptotics of the level counting
function~\cite{gutz},
$\#\{i:E_i\le E\} \sim c_d \vO E^{d/2}$, it follows that
the limit $E\rightarrow\infty$ is equivalent to
$i,j\rightarrow\infty$.
It also follows that for all $d\ge 2$, a growing
number of levels $N_0 \sim c_d\,\vO E^{(d-1)/2}$
falls within an energy window $\ep_0 \sim E^{1/2}$.
This means that, since $\beta$ can take any value strictly below 1/2 that
the order $N$ of the matrix $\tilde{Q}$
which tends to the identity
can be allowed to grow without
limit like $N = o(N_0)$.

We now focus on the special case of Dirichlet boundary conditions,
\be
	\phi_j(\Gamma) \; = \; 0,
\label{eq:dir}
\ee
corresponding to $\Gamma_D = \Gamma$.
In this case a simple calculation shows that the
quadratic form (\ref{eq:qdef}) simplifies such that (\ref{eq:q}) becomes
\be
	Q_{ij} \; = \;
	\oint_\Gamma \! r_n \,\partial_n \phi_i \partial_n \phi_j \, \ds,
\label{eq:qdir}
\ee
that is, a boundary inner product under a weighting function $r_n$.
In this case Corollary~\ref{cor:q} can then be restated as
\begin{cor}[Dirichlet quasi-orthogonality] 
As $E\to\infty$ the
eigenfunction normal derivatives
belonging to an energy window of width $o(E^{1/2})$
become asymptotically orthogonal
under the boundary weighting function $r_n$.
\label{cor:qdir}
\end{cor} 

\begin{figure}
\bc
\includegraphics[width=\textwidth]{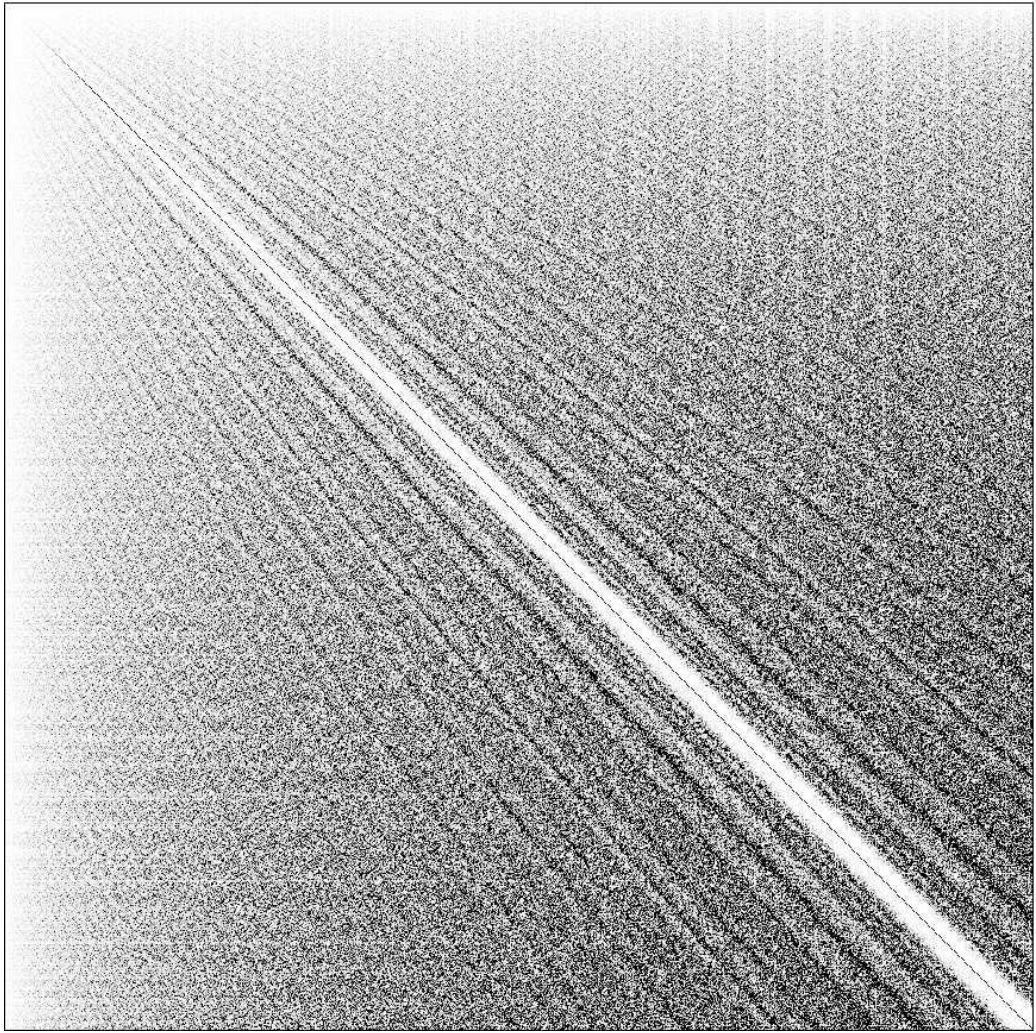}
\ec
\caption{
Density plot of the matrix $Q_{ij}$ for the range $i,j = 1\cdots N$
for $N=1000$.
$i=j=1$ is in the top left corner.
The domain used was a smooth deformation of the
circle, in $d=2$, with Dirichlet boundary conditions;
the classical flow is almost completely ergodic.
By Weyl's Law the energy level density is constant,
therefore the axes can equally well be thought of
as $E_i$ and $E_j$.
Eigenfunctions were calculated using the scaling method
(\cite{que}, Appendix~B).
Dark indicates large magnitude values of $Q$, white indicates zero.
The visible white band (very small values) with growing width $E^{1/2}$
around the diagonal (dark line) shows the quasi-orthogonality property
special to the choice of (\ref{eq:qdef}).
The visibly `banded' structure is discussed in Section~\ref{sec:erg}.
\label{fig:Q}
}
\end{figure}

This Corollary is a stronger form
of a conjecture
first made (we believe\footnote{
Although it was known~\cite{b+w}
to be exact for the degenerate case $E_i=E_j$.})
by Vergini-Saraceno
(see Appendix of \cite{v+s}).
Making use of an identity equating (\ref{eq:qdir}) with
$(E_j-E_i)\langle\phi_i,\mbf{r}\cdot\nabla\phi_j\rangle_\Omega$
for $i\neq j$
(\eg see \cite{mythesis} Eq.(H.25)),
they suggested that off-diagonal elements vanish linearly as $|E_i-E_j|$.
We now know by Theorem~\ref{thm:q}
that the vanishing is of higher order, namely as $(E_i-E_j)^2$.
Understanding this order of vanishing is important for understanding the
accuracy of the numerical `scaling method', as we will discuss below.
%

For domains whose geodesic flow (equivalently, classical dynamics)
is ergodic, a semiclassical
argument (of the type we will discuss in Section~\ref{sec:erg}),
more powerful than that of Vergini-Saraceno,
was developed by Cohen, Heller and
this author.
This predicted the $(E_i-E_j)^2$ factor appearing in (\ref{eq:bound}), and
showed excellent agreement with numerical studies~\cite{dil,mythesis}.
The argument involved evaluation along classical trajectories of a
non-smooth operator which `lives' (is a delta-function)
on the boundary and possesses a `strength' $D\ofs$ (see~\cite{mythesis}).
For the case $D\ofs = r_n$, corresponding to dilation of the billiard,
the operator can be shown to be an exact second time-derivative of
a bounded function.
As a result its power-spectrum vanishes at zero frequency
and this, via the semiclassical argument, is seen to be the cause
of the $(E_i-E_j)^2$ factor.
However this explanation has problems:
i) it relies on a semiclassical argument not yet proven to hold for
all matrix elements
(see Section~\ref{sec:erg}), ii) the
semiclassical argument is not known to be valid for
such singular operators living on the boundary,
and iii) only ergodic domains can be handled.

Corollary~\ref{cor:qdir} provides proof of Dirichlet
quasi-orthogonality {\em without resort to any assumptions about the shape
or ergodicity of the domain}.
In effect the identity in Lemma~\ref{lem:amazing} below bypasses 
the two classical time-derivatives, and the semiclassical argument,
by directly accounting for the $(E_i-E_j)^2$ factor.
Thus quasi-orthogonality is seen to be a {\em result independent of
any semiclassical or quantum ergodic assumptions}.
What we will be left with is consideration of matrix elements of the
operator $r^2$, which is a smooth and bounded function of space.
Bounds on these matrix elements are much easier to find
than in the case of matrix elements involving boundary information.
Lemma~\ref{lem:amazing} relies on overlap identities which follow from
the Divergence Theorem alone;
these identities are derived in the Appendix using a symbolic matrix technique.
Theorem~\ref{thm:q} will then
follow from an elementary bound on matrix elements.
The possibility of considering boundary quasi-orthogonality
with general boundary conditions, as we do with Theorem~\ref{thm:q},
is to our knowledge new.

In terms of applications,
Dirichlet quasi-orthogonality (Corollary~\ref{cor:qdir})
is a key component of the `scaling method' invented by
Vergini-Saraceno~\cite{v+s}
(see \cite{mythesis,scaling}, and Appendix~B of the companion
paper~\cite{que} for review), a very
efficient method for the numerical solution of the Dirichlet eigenproblem
in strictly star-shaped domains ($r_n>0$).
This method has enabled large-scale studies of eigenfunctions at
extremely high $E$ in $d = 2$~\cite{scalinguses,que} and
$d = 3$~\cite{prosen3d}, where it outperforms all other known methods
by a factor of typically $10^3$.
The role played by the matrix $Q$ is as follows:
The small size of off-diagonal elements of $Q$ near the diagonal means
that a set of the domain's
Dirichlet eigenfunctions approximately diagonalize two certain
quadratic forms (see Appendix~B of \cite{que}).
The numerical method involves simultaneous diagonalization of these
two quadratic forms, in order to compute a large number of approximate
domain eigenfunctions at once.
The error in the approximation, and therefore in the functions found
by the method, depends on among
other things the order of vanishing of $Q$ as one approaches the diagonal.
Thus our work finally places this important method on a more rigorous footing,
one which, despite its successful use, it has not yet had.
In particular its success in domains with mixed dynamics (divided phase space),
quasi-integrable, or integrable dynamics
is now no longer mysterious.
Moreover the generality of (\ref{eq:dbcs}) and (\ref{eq:bcs})
opens up the possibility of
extending the scaling method to solve eigenproblems with more general
boundary conditions.

As another application, in the Dirichlet case (\ref{eq:thm1diag}) becomes
\be
	\oint_\Gamma \! r_n (\partial_n \phi_i)^2\, \ds
	\; = \; 2E_i, \qquad \mbox{for all } i\ge 1,
\label{eq:bw84}
\ee
a result first found by Rellich~\cite{rellich} and independently by
Berry-Wilkinson~\cite{b+w}. (Other more simple
derivations have recently been found~\cite{dil,mythesis,hassell}).
This formula is useful numerically because it enables Dirichlet
eigenfunctions to be correctly normalized using boundary information alone.
For general boundary conditions (\ref{eq:thm1diag}) reads
\be
	\oint_\Gamma \! (d-2) \phi_i \partial_n \phi_i
	+ r_n \left(E_i \phi_i^2 - |\nabla \phi_i|^2\right)
	+ 2\partial_r \phi_i \partial_n \phi_i \, \ds
	 =  2E_i, \quad \mbox{for all } i\ge 1,
\label{eq:gen_bw84}
\ee
a normalization formula involving boundary
values and first derivatives alone, which we have not found in the literature.
By contrast, the
general boundary condition
formula of Boasman for $d = 2$ (Appendix~D of \cite{boasman}) requires
second derivatives, a numerically more complicated demand.

Finally we note some other recent work on eigenfunctions on the boundary.
In the Dirichlet case the asymptotic completeness in $C^{\infty}(\Gamma)$
of the boundary functions
has been shown (this holds even when restricted to
an energy window), and the curvature correction to their local
intensity derived and tested numerically~\cite{backerbdry}.
A version of the Quantum Ergodicity Theorem~\cite{qet}
for boundary functions has been proved in
piecewise-smooth domains~\cite{hassellqet}.
We note that boundary normalization formulae such as (\ref{eq:bw84})
break down for general Riemannian manifolds with boundary~\cite{hassell},
so we expect that our results are specific to the Euclidean case.

The remainder of the paper is structured as follows.
In Section~\ref{sec:proof} we prove Theorem~\ref{thm:q}, and
give a recipe for translating the origin in such a way that the bound
$C$ achieves its minimum value $C_0$.
In Section~\ref{sec:que}
we review some semiclassical results on the size of
matrix elements of well-behaved operators as $E\rightarrow\infty$.
Applying them to the operator $r^2$, for domains with ergodic flow
we improve somewhat the
convergence rate in part 2) of Corollary~\ref{cor:q}.
However, in contrast to our main results,
this improved ergodic statement is not proved to hold for
every single element $i,j$.
We will also briefly mention expected behaviour for
domains with integrable and mixed flow.
The Appendix contains a detailed account of a general
symbolic matrix procedure
used to derive certain identies used in the proof in Section~\ref{sec:proof}.

{\bf Acknowledgements}.
This work could not have been possible without interactions with and important
feedback from Percy Deift, Fanghua Lin,
Peter Sarnak, Maciej Zworski, Kevin Lin, Doron Cohen and Eduardo Vergini.
The boundary formulae of Appendix~A originated in work with Michael Haggerty
and Eric Heller.
The author is supported by the Courant Institute at New York University.

\section{Proof of Theorem~\ref{thm:q}}
\label{sec:proof}

Theorem~\ref{thm:q} hinges on
the following identity between elements of the matrix $Q$
and matrix elements of the function $r^2$ over the domain.
\begin{lem} 
Let $Q$ be the matrix defined by (\ref{eq:qdef}) and (\ref{eq:q}),
with Laplace eigenfunctions defined by (\ref{eq:helm}), (\ref{eq:dbcs})
and (\ref{eq:bcs}). Then,
\be
	Q_{ij} \; = \; 
	2E_i \delta_{ij} \;+\; 
	\frac{\epij^2}{4}\rrij, \qquad \mbox{for all } i,j \ge 1,
\label{eq:amazing}
\ee
where $\epij\equiv E_i-E_j$ is the difference in energies.
\label{lem:amazing}
\end{lem} 

Notice that in (\ref{eq:amazing}) the first (resp.\ second) term
contributes only on (resp. off) the diagonal.

{\bf Proof.}
Consider functions $u$ and $v$,
defined with their derivatives on $\overline{\Omega}$, and each satisfying
the Helmholtz equation, namely,
\bea
	-\Delta u &= & E_u u, \nonumber \\
	-\Delta v &= & E_v v,
\label{eq:helmuv}
\eea
everywhere in $\Omega$.
The functions need satisfy no particular boundary conditions on $\Gamma$.
There are two cases which are handled separately: the
energies $E_u$ and $E_v$ are either equal or unequal.
Examining first the equal-energy case $E_u = E_v = E$,
in the Appendix we derive the identity (see (\ref{eq:overlap_eq})),
\be
	\uv
	\; = \;
	\frac{1}{2E} \oint_{\Gamma}\!
	\frac{d-2}{2}(u_n v+v_n u)
	+ r_n\left(E uv - \nabla u \cdot \nabla v\right)
	+ u_r v_n + v_r u_n \, \ds.
\label{eq:same_gen}
\ee
This identity is a consequence of the Divergence Theorem applied to
various combinations of the functions $u$ and $v$ and their first derivatives.
Comparing this with (\ref{eq:qdef}) gives
\be
	\uv \; = \frac{1}{2E}q(u,v;2E).
\ee
Substituting $u=\phi_i$ and $v=\phi_j$ and using the (\ref{eq:q})
and orthonormality on the domain gives,
\be
	Q_{ij} \; = \; 2E_i \delta_{ij},
	\qquad \mbox{for} \; E_i=E_j.
\label{eq:same_q}
\ee
This covers both the diagonal element case $i=j$ and the
degenerate case where $i\neq j$ but $E_i=E_j$.

Now examining the case $E_u \neq E_v$,
in the Appendix we derive the identity (equivalent to (\ref{eq:overlap_uneq})),
\begin{multline}
	\frac{\ep^2}{4}\rruv
	\; = \;
	\oint_{\Gamma}\!
	\frac{d-2}{2}(u_n v+v_n u)
	+ \left(\frac{\et}{\ep} - \frac{\ep}{4}r^2\right)(u_n v-v_n u)\\
	+ r_n\left(\frac{\et}{2} uv - \nabla u \cdot \nabla v\right)
	+ u_r v_n + v_r u_n \, \ds,
\label{eq:diff_gen}
\end{multline}
where the energy difference is $\ep \equiv E_u-E_v$ and the sum
$\et\equiv E_u+E_v$.
Applying self-adjoint
boundary conditions (\ref{eq:dbcs}) and (\ref{eq:bcs})
causes the term anti-symmetric
in $u$ and $v$ to vanish identically on the boundary, leaving
\be
	\frac{\ep^2}{4}\rruv \; = \;
	q(u,v; \,E_u+E_v) .
\ee
Substituting $u=\phi_i$ and $v=\phi_j$ and using (\ref{eq:q}) gives
\be
	Q_{ij} \; = \; \frac{\epij^2}{4}\rrij
	\qquad \mbox{for} \;E_i\neq E_j.
\label{eq:diff_q}
\ee
The two cases---equal energy (\ref{eq:same_q}) and
differing energy (\ref{eq:diff_q})---can be summarized
by the one identity (\ref{eq:amazing}).
\hfill $\square$

\vspace{1ex}

We now see that the particular form (\ref{eq:qdef}) of $q$ results
from a relatively complicated sequence of manipulations detailed in
the Appendix.
Lemma~\ref{lem:amazing} has reduced the question of quasi-orthogonality to
one of the size of the matrix elements $\rrij$.
Now the diagonal elements $\rrii$ are trivially bounded by
\be
	\int_\Omega\! r^2 \phi_i^2\,\dr
	\; \le \;
	R^2 \int_\Omega \! \phi_i^2 \,\dr
	\; = \; R^2,
\ee
where $\dr$ is the volume element, and
the maximum radius from the origin attained by the domain is
\be
	R \; \equiv \; \max_{\mbf{r}\in\Gamma} |\mbf{r}|.
\ee
Then by Cauchy-Schwarz we have off-diagonal matrix elements bounded
by the same constant,
\be
	|\rrij| \; \le \; |\rrii|^{1/2} |\rrjj|^{1/2}
	\; \le \;
	R^2.
\ee
Substituting this into (\ref{eq:amazing}) completes the proof of
Theorem~\ref{thm:q}, with the constant being $C = R^2/4$.
\hfill $\square$

\vspace{1ex}

Clearly the eigenproblem (\ref{eq:helm}), (\ref{eq:dbcs}) and (\ref{eq:bcs}) is
translationally invariant, that is, $\{E_i\}$ and $\{\phi_i\}$
are independent
of the choice of origin for our coordinate $\mbf{r}$.
It follows from (\ref{eq:thm1diag}) that the diagonal elements
of $Q$ are also translationally invariant;
however the off-diagonal elements of $Q$ are not.
Let us assume our task
is, given a domain $\Omega$ of particular shape, to find the
boundary form given by (\ref{eq:qdef})
for which the constant in (\ref{eq:bound}) can be chosen to be smallest.
We are free to translate the origin to achieve this goal.
Clearly the minimum $C$ is given by $C_0 = R_0^2/4$, where $R_0$ is
the radius of the smallest circle (generally, ball in $\mathbb{R}^d$)
enclosing $\Omega$. This is the escribed circle (or ball); see Fig.~\ref{fig}b.
The optimal choice of origin is the center of this circle, $\mbf{r}_0$.
We note that any choice for the origin
need not fall inside $\Omega$ and that none of the results given in this
paper depend on whether it does or not.
(This contrasts the scaling method, for which it is believed that the
domain must be strictly star-shaped with respect to the origin).

\section{Semiclassical estimates}
\label{sec:que}

The power of
Lemma~\ref{lem:amazing} is that it has reduced the quasi-orthogonality question
to one of the size of matrix elements
of a bounded operator $r^2$,
in such a way that a trivial bound on $\rrij$
is adequate to prove Theorem~\ref{thm:q}.
Can we go beyond this bound?
Estimating matrix elements semiclassically (at large $E$)
has been a major theme
in both the physics and mathematics
communities, and remains an active area of research.
We will now draw on their results. We emphasize that to prove
Theorem~\ref{thm:q}, semiclassical results have not been used.

Depending on the domain (`billiard')
shape, the geodesic flow on $\Omega \times S^{d-1}$,
that is, the classical motion of a trapped point particle undergoing
elastic reflections from $\Gamma$,
falls into three broad categories: ergodic, integrable, and
mixed~\cite{ott}.
This has consequences for the Laplace eigenfunctions which have been
a major theme in `quantum chaos'~\cite{gutz,sarnak}.

\subsection{Ergodic domains}
\label{sec:erg}

In the ergodic case, the Quantum Ergodicity Theorem~\cite{qet} (QET) has been
proved, stating that asymptotically ($E\rightarrow\infty$)
all but a set of measure zero
of the eigenfunctions become spatially
equidistributed across the domain.
Thus almost all diagonal elements $\rrii$ tend to the classical
average $\int_\Omega r^2 \, \dr/\vO$.
However this is not of direct use since
it is only {\em off-diagonal} elements of $r^2$
that play a role in
Lemma~\ref{lem:amazing}.
Assuming a statistical model in which
eigenfunctions behave like uncorrelated random-waves~\cite{berry77},
off-diagonal elements are expected
to be Gaussian distributed with zero mean; this has indeed been verified
numerically~\cite{austin,prosen}.

The remaining issue is their variance. This issue is discussed in
much more detail in \cite{que} but we provide a summary here.
Almost all off-diagonal elements have been proven to vanish
asymptotically~\cite{zeloffdiag} implying that the variance
must vanish. However, the random-wave
prediction for variance size is poor~\cite{austin,dil,que}
since it cannot account for variance changing as a function of energy
difference, that is, a `banded structure' to the matrix.
The banded matrix structure in the case of an ergodic domain is shown in
Fig.~\ref{fig:Q}.
The semiclassical sum rule of
Feingold-Peres~\cite{fp86,EFKAMM} (FP) has proven much more successful
in predicting numerically observed
variance~\cite{fp86,EFKAMM,austin,prosen,dil},
including the banded structure.
There is numerical evidence~\cite{que}
that it is asymptotically correct in
billiard systems, but that convergence is quite slow.
The band profile of the matrix elements of an operator $\hat{A}$,
that is, variance as a function of energy difference,
is related to $\tilde{C}_A(\omega)$,
the Fourier transform
of the auto-correlation of the corresponding classical operator under
the (unit-speed) classical flow.
FP states that for operators $\hat{A}$ whose classical
counterparts are well-behaved in phase-space,
\be
	\mbox{var}\left[\langle \phi_i, \hat{A}\phi_j\rangle_\Omega\right]
	\; \rightarrow \;
	\frac{\tilde{C}_A(\omega_{ij})}{\vO} E^{-1/2}
	\qquad \mbox{for } E_i \approx E_j \approx
	E\rightarrow \infty,
\label{eq:fp}
\ee
where the wavenumber difference is
$\omega_{ij} \equiv E_i^{1/2} - E_j^{1/2}$.
The sense in which we use `variance' here is the
variance of a large
sample of matrix elements which share similar values of $\omega_{ij}$
and fall within a classically-small energy range.
Thus, we are measuring variance within a small `patch' of the matrix
$\langle \phi_i, \hat{A}\phi_j\rangle_\Omega$.
As with Corollary~\ref{cor:q}, we now restrict
ourselves to the energy window $\ep_0 = o(E^{1/2})$,
within which $\omega_{ij} \rightarrow 0$ as $E\rightarrow\infty$
(corresponding classically to taking the zero-frequency limit).
Choosing $\hat{A} = r^2$, then (\ref{eq:fp}) and Lemma~\ref{lem:amazing}
gives matrix element size
\be
	|Q_{ij}| \; \approx \;
	\left(\mbox{var}\left[Q_{ij}\right]\right)^{1/2} \; \rightarrow \;
	\left(\frac{\tilde{C}_{r^2}(0)}{4\vO}\right)^{1/2} \!\!\!
	E^{-1/4} \,(E_i-E_j)^2,
\label{eq:ergasym}
\ee
which holds for nearly all off-diagonal elements,
and should be compared to the hard bound (\ref{eq:bound}).
The domain-dependent constant $\tilde{C}_{r^2}(0)$
can be estimated numerically via trajectory simulations;
we do not know if it is minimized by the optimal
choice of origin $\mbf{r}_0$ derived in Section~\ref{sec:proof}.
Thus in part 2) of Corollary~\ref{cor:q}, we can improve the
bound to $c_\beta E^{2\beta - 5/4}$.
For $\beta=0$ (constant energy difference), the convergence
rate then improves somewhat from $E^{-1}$ to $E^{-5/4}$.
Note that this result is equivalent to that of \cite{dil,mythesis},
but because $r^2$ is a smooth rather than singular
function the approach presented here stands on more solid foundations,
as discussed in Section~\ref{sec:main}.
However, FP is not proven to hold for every single matrix element:
there may
exist an exceptional set (analogous to the diagonal case) which
fails to converge as above. Theorem~\ref{thm:q} remains the only
proven bound on $Q$ known to us.

It is worth pointing out that ergodicity and the FP sum rule,
if it were valid for singular functions (as it empirically
appears to be~\cite{dil}),
would imply a
weaker form of asymptotic quasi-orthogonality,
regardless of the form of $q$. That is, Corollary~\ref{cor:q}
would hold even
for {\em generic} boundary forms with terms similar to those in (\ref{eq:qdef})
but lacking their particular cancellations,
for almost all off-diagonal elements, independent of $E_i-E_j$,
but with the much slower convergence rate $\sim cE^{-1/4}$.
In the case of Dirichlet boundary conditions, this would correspond to
choosing a generic weight function $D\ofs$ in place of $r_n$
in (\ref{eq:qdir}).

When the ergodic flow furthermore is Anosov (uniformly hyperbolic), a
stronger version of QET has been conjectured~\cite{sarnak}, which lifts
the requirement that a zero-density set be excluded.
This is known as Quantum Unique Ergodicity (QUE), and currently
remains unproven in Euclidean domains, although good numerical evidence
exists when $d = 2$ \cite{que}.
Since QUE can be extended to off-diagonal elements~\cite{zeloffdiagque},
if proven, QUE would enable the improved bounds (\ref{eq:ergasym})
to be claimed for every single off-diagonal element in these classes of
domains.

If the flow contains orbits with zero Lyapunov exponent (`bouncing ball
modes'), classical autocorrelations generally have a power-law rather
than exponential tail,
thus $\tilde{C}_A(\omega)$ diverges as $\omega\rightarrow 0$,
and (\ref{eq:ergasym}) is undefined.
In this case FP must be modified, and to our knowledge only the
diagonal variance has been carefully studied theoretically~\cite{EFKAMM}.

\subsection{Integrable and mixed domains}

Finally we briefly mention some of what is known about the behaviour of matrix
elements in systems with other categories of classical dynamics,
and speculate on consequences for the rate of convergence
in Corollary~\ref{cor:q}.

If the classical dynamics is integrable, then eigenfunctions are regular,
with Wigner functions
concentrating around $d$-dimensional classical invariant
tori in phase space~\cite{ott,gutz}. Each eigenfunction possesses a
well-defined set of quantum number labels $\mbf{p} \in \mathbb{Z}^d$;
values of $\mbf{r}$ tend to be uncorrelated as a function of level number.
Much less has been said about off-diagonal matrix elements in this case.
In smooth Hamiltonian systems matrix elements between
levels labelled by $\mbf{p}$ and $\mbf{q}$ are expected to die like
$\exp(-c|\mbf{p}-\mbf{q}|)$,
for a generic
smooth operator. Thus most off-diagonal elements are exponentially small
(the so-called `selection rules'~\cite{gutz,mehlig}), but a few may be
large, that is, of the same order as diagonal elements.
However, because our domain has a hard boundary condition,
it is easy to estimate (by analogy with the $d = 1$ case) that generic
matrix elements die algebraically with $|\mbf{p}-\mbf{q}|)$.
Thus we still expect most off-diagonal elements to be small.
This has been observed numerically in billiard systems~\cite{prosen,mehlig}.
Thus, we expect that convergence to quasi-orthogonality will
be much more rapid, for the majority of matrix elements, than for
the ergodic case. In general, it is hard to say more.
It would be interesting to see if integrable examples can be found
where $\rrij$ approaches $R^2$ as $E\rightarrow\infty$;
this would prove that
Theorem~\ref{thm:q} is sharp, with $C = R^2/4$.

A generic domain has mixed dynamics, and eigenfunctions either occupy
ergodic components of phase space, or lie on integrable tori.
Matrix elements involving energy levels lying in different components
are therefore expected to be very small.
This has been verified
numerically in some smooth Hamiltonian systems~\cite{boose}, but no study
in billiards is known. There is much opportunity for
numerical study in such billiard systems.

\section*{Appendix A: Boundary identities for matrix elements}
\label{sec:bdry}

We present and apply a novel general procedure for deriving
identities which express certain bilinear forms over a domain, involving
solutions to the Helmholtz equation
purely in terms of bilinear
forms on the boundary.
Given a bounded open domain $\Omega \subset \mathbb{R}^d$, $d\ge 2$,
let the fields $u\ofr$, $v\ofr$ be regular solutions of
(\ref{eq:helmuv}) with energies
$E_u$ and $E_v$
respectively, everywhere in $\Omega$.
We envisage $\Gamma$ being the union of a finite number of smooth curves
(surfaces),
however the broader condition that $\Omega$ be
Lipschitz is sufficient.
Crucially, we do not require any particular boundary conditions to be
satisfied by $u$, $v$ on $\Gamma$.
We will make use of the Divergence Theorem applied to vector fields which are
bilinear combinations of $u, v$ and their first derivatives; thus
$u$ and $v$ need only be sufficiently regular that the Divergence
Theorem can be applied to such a vector field. These conditions on
vector fields are quite broad~\cite{kellogg}.
We believe that the conditions laid out at the beginning of this paper
are sufficient to ensure that $u$ or $v$ can be set to equal any
eigenfunction $\phi_i$ and still have the Divergence Theorem apply.
We will express
\be
	\langle u, \hat{H}v\rangle_{\Omega} := \int_{\Omega}
	\! u\hat{H}v\,\dr,
\ee
$\dr$ being the volume element in $\mathbb{R}^d$,
in terms of boundary integrals involving values and first derivatives
of $u$ and $v$, for the operator choices $\hat{H} = 1$ and
$\hat{H} = \mbf{r}\cdot\mbf{r} := r^2$.
We will consider both the cases $E_u\neq E_v$ and $E_u=E_v$.

Consider functions on $\Omega$ that are
built from bilinear forms in $u, v$ and their spatial derivatives.
We restrict attention to such functions that are either
scalar fields or vector fields.
The key is to
select a set of scalars $\{q_\beta\}$, $\beta = 1,2\ldots \nq$,
and a set of vectors $\{\mbf{p}_\alpha\}$,
$\alpha = 1,2\ldots \np$,
such that the divergence of
each vector field can be written as a (location-independent)
linear combination of the scalar fields.
Applying the Divergence Theorem turns these into relations between
boundary and volume integrals.
The coefficients form a matrix, which can be handled symbolically to express
volume integrals as linear combinations of boundary integrals.
We note that the idea of inverting such a coefficient matrix
originated in unpublished work of
Michael Haggerty (see Appendix H of \cite{mythesis}),
however here we extend the idea and
explore a larger set of vector and scalar terms.

\begin{table}
\bc
\begin{tabular}{l|l|l|l|}
order& $r^0$& $r^1$ & $r^2$ \\
\hline
$\partial^0$& $uv$ & --- & $\la{rr}uv$ \\
\hline
$\partial^1$& $\la{uv}$ & $\la{ru}v$, \lr &
	$\la{rr}\la{uv}$, $\la{ru}\la{rv}$\\
\hline
$\partial^2$& $\di{uv}{\lla}$ & $\di{ruv}{\lc\lc}$, \lr &
	$\la{rr}\di{uv}{\lla}$, $\di{ruvr}{\lc\lc\lc}$,
	$\di{rur}{\lc\lc}v$, \lr\\
\hline
\end{tabular}
\ec
\caption{Lowest order scalar diagrams (`molecules') from which the set
of functions
$\{q_\beta\}$ was chosen. An order of $\partial^m$ means that
$m$ is the highest order derivative of $u$ or $v$.
The symbol \lr\ means a repetition of the previous
diagram but with $u$ and $v$ swapped. See text for further explanation.
\label{tbl:sca}
}
\end{table}
\begin{table}
\bc
\begin{tabular}{l|l|l|l|}
order& $r^0$& $r^1$ & $r^2$ \\
\hline
$\partial^0$& --- & $\da{r}uv$ & ---\\
\hline
$\partial^1$& $\da{u}v$, \lr &
	$\da{r}\la{uv}$, $\la{ru}\da{v}$, \lr  &
	$\la{rr}\da{u}v$, \lr, $\da{r}\la{ru}v$, \lr\\
\hline
\end{tabular}
\ec
\caption{Lowest order vector diagrams (`radicals') from which the set
of functions
$\{\mbf{p}_\alpha\}$ was chosen. See Table~\ref{tbl:sca} caption
for explanation.
\label{tbl:vec}
}
\end{table}

To build the functions we use the following objects:
the fields $u,v$, their gradients $u_i := \partial_i u$
and higher derivative tensors
$u_{ij} := \partial_i \partial_j u$, etc, and
polynomials in the coordinates $r_i$ which transform as scalars,
vectors, tensors, etc.
It is possible to enumerate systematically, using diagrams, the
hierarchy of such combinations with overall scalar or vector character.
Diagrams also greatly ease the algebra in taking the divergence
of the vector functions (we apply the rules by hand, but
automating them would not be hard).

We have found the following scheme very useful.
A combination of overall scalar character is represented by a `molecule'
(or a disconnected collection of molecules)
built from `atoms' connected by `bonds'. Each atom can be either $r$, $u$,
or $v$. A bond carries summation over a spatial index, for instance $i$;
when a bond
terminates at $u$ or $v$ this corresponds to a derivative $\partial_i$,
whereas termination at $r$ corresponds simply to $r_i$.
Thus $u$ and $v$ may carry $0,1,\ldots\infty$ bonds whereas each $r$ must
carry exactly 1. See Table~\ref{tbl:sca} for the few lowest-order
scalar functions.
For example, $\di{ruv}{\lc\lc}$ represents
$\sum_{i=1}^d \sum_{j=1}^d r_i u_{ij} v_j$. From now on summation
over repeated indices is implied.

A combination of overall vector character is similar
except it has exactly one `dangling bond', represented
by a dot (we could think of it as a `radical').
Taking the divergence of a vector function
corresponds to first summing the molecules resulting
from connecting the dangling bond to each of the atoms in turn (including
the one from which it originates). The molecules produced are not
necessarily valid, so the following three `reaction' rules need to be
applied repeatedly until each molecule becomes valid:
\ben
\item	$\di{u}{\ls} \rightarrow \mku u$. That is, self-bonds to atoms
$u$ or $v$ vanish and are replaced by the scalar multiplier $\mku$
or $\mkv$ respectively.
\item	$\di{r}{\ls} \rightarrow d$. That is,
any $r$ bonded to itself vanishes along with the bond,
and is replaced by the scalar multiplier $d$.
\item	$\di{r}{\lc\lc} \rightarrow \la{\rule{0ex}{1ex}}$.
That is, any $r$ with two bonds (not covered by the previous rule)
vanishes, and the resulting dangling ends connect to form a single bond.
\een
It can be easily verified that this procedure is equivalent to taking
the divergence of any given vector field.
For example, the divergence of $r^2 u_i v$ is computed as follows,
\bea
	\la{rr}\da{u}v & \rightarrow &
	2\di{rru}{\lc\lc}v + \la{rr}\di{u}{\ls}v + \la{rr}\la{uv}
	\nonumber\\
	& \stackrel{\tbox{rules}}{\rightarrow} &
	2\la{ru}v - E_u\la{rr}uv + \la{rr}\la{uv},
\eea
thus the answer is $2r_i u_i v - E_u r^2 uv + r^2 u_i v_i$,
evident as row $\alpha=7$ of the matrix equation (\ref{eq:mat8}) below.

Using this technique we have calculated the divergence of the
simplest 25 or so vectors, which results in a similar number of scalars.
For the goal at hand we have eventually found the following vector subset
of size $\np=8$ and scalar subset of size $\nq = 8$
to be minimally sufficient,
\be
\partial_i
	\left(\begin{array}{c}
	u_i v\\
	v_i u\\
	r_i uv\\
	r_i u_j v_j\\
	r_j u_j v_i\\
	r_j v_j u_i\\
	r^2 u_i v\\
	r^2 v_i u\\
	\end{array} \right)
\;\;=\;\;
	\left(\begin{array}{cccccccc}
	\mku&1&0&0&0&0&0&0\\
	\mkv&1&0&0&0&0&0&0\\
	d&0&1&1&0&0&0&0\\
	0&d&0&0&1&1&0&0\\
	0&1&\mkv&0&1&0&0&0\\
	0&1&0&\mku&0&1&0&0\\
	0&0&2&0&0&0&\mku&1\\
	0&0&0&2&0&0&\mkv&1
	\end{array}\right)
	\left(\begin{array}{c}
	uv\\
	u_i v_i\\
	r_i u_i v\\
	r_i v_i u\\
	r_i u_{ij} v_j\\
	r_i v_{ij} u_j\\
	r^2 uv\\
	r^2 u_i v_j
	\end{array} \right).
\label{eq:mat8}
\ee
This set of divergence relations can be more compactly written,
\be
	\nabla \cdot \mbf{p}_\alpha \; = \; \sum_{\beta=1}^{\nq}
	{\cal M}_{\alpha \beta} \,q_\beta , \qquad \alpha = 1,2\ldots \np.
\label{eq:matrix}
\ee
Integrating over $\Omega$ and applying the Divergence Theorem gives
\be
	\oint_{\Gamma}\!\mbf{n} \cdot \mbf{p}_\alpha \ds
	\; = \;
	\sum_{\beta=1}^{\nq}
	{\cal M}_{\alpha \beta} \, \int_{\Omega} \!q_\beta \,\dr,
	\qquad \alpha = 1,2\ldots \np.
\label{eq:vol}
\ee

We first consider the case $E_u \neq E_v$, for which
the matrix ${\cal M}$ is invertible (in fact we chose the above
set of vectors to make this so). Thus each
volume integral appearing in (\ref{eq:vol}) can be written purely in terms of
boundary integrals,
\be
	\int_{\Omega} \!q_\alpha \,\dr
	\; = \;
	\sum_{\beta=1}^{\np}
	\left({\cal M}^{-1}\right)_{\alpha \beta}
	\oint_{\Gamma}\!\mbf{n} \cdot \mbf{p}_\beta \,\ds
	\qquad \alpha = 1,2\ldots \nq. 
\label{eq:invert}
\ee
A symbolic algebra package (Mathematica) was used to find
${\cal M}^{-1}$, which we need not write out in full here.
Rather we use only select rows of ${\cal M}^{-1}$, that is, certain
values of $\alpha$ in (\ref{eq:invert}).
For example, the $\alpha=1$ row of ${\cal M}^{-1}$ is
$(1/(E_u-E_v),-1/(E_u-E_v),0,0,0,0,0,0)$,
immediately gives the simple relation
\be
	\langle u,v \rangle_{\Omega_A} \; = \; \frac{1}{E_u - E_v}
	\oint_{\Gamma_A} \! (u \mbf{n}\cdot\nabla v - v \mbf{n}\cdot\nabla u)
	\; \ds,
\label{eq:ovgreen}
\ee
well-known from Green's Theorem.
Choosing the $\alpha=7$ row gives,
expressed in terms of $\ep:= E_u-E_v$ and $\et:= E_u+E_v$,
the relation
\begin{multline}
	\int_{\Omega}\!r^2 uv\,\dr \; = \;
	\frac{4}{\ep^2}\oint_{\Gamma}\!
	\frac{d-2}{2}(u_n v+v_n u)
	+ \left(\frac{\et}{\ep} - \frac{\ep}{4}r^2\right)(u_n v-v_n u)\\
	+ r_n\left(\frac{\et}{2} uv - \nabla u \cdot \nabla v\right)
	+ u_r v_n + v_r u_n \, \ds.
\label{eq:overlap_uneq}
\end{multline}
In the particular case of Dirichlet boundary conditions on $\Gamma$,
using $\nabla u = \mbf{n}u_n$, etc,
this relation simplifies to
\be
	\int_{\Omega}\!r^2 uv\,\dr \; = \;
	\frac{4}{\ep^2}\oint_{\Gamma}\!r_n u_n v_n \ds,
\ee
a result we have not found in the literature.

We now consider $E_u = E_v = E$, in which case ${\cal M}$ is not invertible
(its first two rows are identical). However $\int_\Omega q_\alpha\,\dr$
can still be expressed as a linear combination of boundary integrals
if the standard unit basis vector $\mbf{b}^{(\alpha)} :=
(0,\ldots,0,1,0,\ldots,0)\in\mathbb{R}^\nq$,
which contains a 1 only in location $\alpha$,
lies in $\mbox{Row} {\cal M}$.
Since in our case
$\mbox{Nul} {\cal M}$ contains the single vector $(0,0,0,0,0,0,1,E)$,
this can be done for $\alpha \in [1,6]$.
For each such $\alpha$, we can
find the row coefficient vector
$\mbf{x}^{(\alpha)}\in\mathbb{R}^\np$
by solving the (consistent and in this case underdetermined) linear equation
\be
	{\cal M}^T \mbf{x}^{(\alpha)}
	\; = \; \mbf{b}^{(\alpha)}.
\label{eq:Msolv}
\ee
We remind the reader that the
coefficient vector spaces between which ${\cal M}^T$ transforms
should not be confused with coordinate vectors in $\mathbb{R}^d$.
The solution set of (\ref{eq:Msolv})
is readily found with a symbolic algebra package.
For the case $\alpha = 1$ the resulting row coefficients give
the identity
\be
	\int_{\Omega}\!uv\,\dr \; = \;
	\frac{1}{2E} \oint_{\Gamma}\!
	\frac{d-2}{2}(u_n v+v_n u)
	+ r_n\left(E uv - \nabla u \cdot \nabla v\right)
	+ u_r v_n + v_r u_n \, \ds.
\label{eq:overlap_eq}
\ee
where we have (arbitrarily) chosen to equalize the first two coefficients
in order to make the $u{\leftrightarrow}v$ symmetry manifest.
This identity
involves no derivatives higher than the first, and to our knowledge is new.
A similar identity for $d=2$ which required second derivatives, and is
therefore more cumbersome, has already been found~\cite{boasman}
(also see \cite{mythesis}, Eq.(H.7)).
In the particular case of Dirichlet boundary conditions (\ref{eq:overlap_eq})
directly proves (\ref{eq:bw84}).

What could be gained by enlarging the function spaces beyond $\np=8$ and
$\nq = 8$ ?
For both cases $E_u=E_v$ and $E_u\neq E_v$, the
only criterion for being able to express
a volume integral $\int_\Omega q_\alpha\,\dr$ in terms of boundary
integrals is that the unit vector $\mbf{b}^{(\alpha)}$ lie in
$\mbox{Row} {\cal M}$. This is
equivalent to the consistency of (\ref{eq:Msolv}).
Invertibility of $\cal M$ is sufficient but not necessary.
It is
an open question
whether, by including a growing set
of vector functions $\{\mbf{p}_\beta\}$ (Table~\ref{tbl:vec} and its
continuation), and necessarily
the resulting growing set of scalars $\{q_\alpha\}$,
that the row space of the resulting growing matrix $\cal M$ can be
made to contain the unit vector $\mbf{b}^{(\alpha)}$ for any desired $\alpha$.

Note that
the methods of this appendix are probably not simply generalizable
to Riemannian manifolds with boundary. The Euclidean
boundary formula (\ref{eq:bw84}) is known to break down in a general
manifold, and unit-normalized eigenfunctions with exponentially-small
values and gradients everywhere on the boundary become
possible~\cite{hassell}.
This is associated with the existence of trapped geodesics.
However for the constant-curvature case, it would be interesting to
search for a generalization.



\end{document}